\documentclass{article}
\usepackage{hiph-art}
\volnumber{19} \issuenumber{1} \edyear{2004}                             
\frompage{000} \topage{000}                                              
\recrevdate{1 January 2004}                                              
\def\rightmark{Nucleon form factors}
\def\leftmark{R. Bijker}
 
\title{The structure of the nucleon} 
\authors{ 
{Roelof Bijker$^1$%
\index{Bijker, R.} 
}\\[2.812mm]
{\normalsize
\hspace*{-8pt}$^1$ Instituto de Ciencias Nucleares,\\ 
Universidad Nacional Aut\'onoma de M\'exico,\\
A.P. 70-543, 04510 M\'exico, D.F., M\'exico
}}
 
\abstract{Experimental data on electromagnetic and weak form factors of the 
nucleon are analyzed in a two-component model with a quark-like intrinsic 
structure surrounded by a meson cloud. The contribution from strange quarks 
is discussed and compared with recent data from the G0 Collaboration.}


 
\makeindex

\newcommand{\ba}{\begin{eqnarray}}
\newcommand{\ea}{\end{eqnarray}}

\begin{document}
 
\maketitle

\section{Introduction}
 
Electromagnetic and weak form factors 
are key ingredients to the understanding of the internal structure 
of the nucleon, since they contain the 
information about the distributions of charge and magnetization. 
Evidence that the nucleon is a composite particle is, among others, 
provided by the anomalous magnetic moment, the finite size, and the 
scaling properties observed in deep-inelastic scattering experiments. 

The complex structure of the proton manifested itself once again in 
recent polarization transfer experiments \cite{jones} which showed 
that the ratio of electric and magnetic form factors of the proton 
exhibits a dramatically different behavior as a function of the momentum 
transfer as compared to the generally accepted picture of form factor 
scaling obtained from the Rosenbluth separation method \cite{andivahis}. 

In recent experiments, parity-violating elastic electron-proton 
scattering has been used to probe the contribution of strange quarks 
to the structure of the nucleon. The strange quark content of the 
form factors can be determined assuming charge symmetry and combining 
parity-violating asymmetries with measurements of the electric and 
magnetic form factors of the proton and neutron. 

The aim of this contribution is to present a study of the electromagnetic 
and strange form factors of the nucleon in a two-component model \cite{IJL,BI}. 
As an illustration, the results for proton and neutron form factor ratios and 
strange form factors are presented and compared to recent experimental data. 

\section{Nucleon form factors} 
 
The form factors of the nucleon arise from matrix elements of the 
corresponding current operators
\ba
\left< N \left| J_{\mu} \right| N \right> = \bar{u}_N \left[ 
F_1(Q^2) \, \gamma^{\mu} + \frac{i}{2M_N} F_2(Q^2) \, \sigma^{\mu\nu} q_{\nu} 
\right] u_N ~.
\ea
Here $F_{1}$ and $F_2$ are the Dirac and Pauli form factors 
which are functions of the squared momentum transfer $Q^2=-q^2$. 
The electric and magnetic form factors, $G_{E}$ and $G_{M}$, are 
obtained from $F_{1}$ and $F_{2}$ by the relations $G_E=F_1-\tau F_2$ 
and $G_M=F_1 + F_2$ with $\tau=Q^2/4 M_N^2$. 

Different models of the nucleon correspond to different assumptions about 
the Dirac and Pauli form factors. In the present model, the external photon 
couples both to an intrinsic three-quark structure described by the form 
factor $g(Q^2)$, and to a meson cloud via vector-meson ($\rho$, $\omega$ 
and $\phi$) dominance (VMD). In the original VMD calculation \cite{IJL}, 
the Dirac form factor was attributed to both the intrinsic structure and 
the meson cloud, and the Pauli form factor entirely to the meson cloud. 
In \cite{BI}, it was shown that the addition of an intrinsic 
part to the isovector Pauli form factor as suggested by studies of 
relativistic constituent quark models in the light-front approach 
\cite{frank,salme}, improves the results for the neutron elecric and magnetic 
form factors considerably. These considerations lead to the following form of 
the isoscalar and isovector form factors \cite{BI}
\ba
F_{1}^{I=0}(Q^{2}) &=& \frac{1}{2} g(Q^{2}) \left[ 
1-\beta_{\omega}-\beta_{\phi} 
+\beta_{\omega} \frac{m_{\omega }^{2}}{m_{\omega }^{2}+Q^{2}} 
+\beta_{\phi} \frac{m_{\phi}^{2}}{m_{\phi }^{2}+Q^{2}}\right] ~, 
\nonumber\\
F_{2}^{I=0}(Q^{2}) &=&\frac{1}{2}g(Q^{2})\left[ 
\alpha_{\omega} \frac{m_{\omega }^{2}}{m_{\omega }^{2}+Q^{2}} 
+ \alpha_{\phi} \frac{m_{\phi}^{2}}{m_{\phi}^{2}+Q^{2}}\right] ~,
\nonumber\\
F_{1}^{I=1}(Q^{2}) &=&\frac{1}{2}g(Q^{2})\left[ 1-\beta_{\rho} 
+\beta_{\rho} \frac{m_{\rho}^{2}}{m_{\rho}^{2}+Q^{2}} \right] ~,  
\nonumber\\
F_{2}^{I=1}(Q^{2}) &=&\frac{1}{2}g(Q^{2})\left[ 
\frac{\mu_{p}-\mu_{n}-1-\alpha_{\rho}}{1+\gamma Q^{2}} 
+\alpha_{\rho} \frac{m_{\rho }^{2}}{m_{\rho}^{2}+Q^{2}} \right] ~.  
\label{ff}
\ea
This parametrization ensures that the three-quark contribution to the anomalous 
magnetic moment is purely isovector, as given by $SU(6)$. For the intrinsic form 
factor a dipole form $g(Q^{2})=(1+\gamma Q^{2})^{-2}$ is used which is 
consistent with p-QCD and coincides with the form used in 
an algebraic treatment of the intrinsic three-quark structure \cite{bijker}. 

The large width of the $\rho$ meson is crucial for the small $Q^{2}$ behavior 
of the form factors and is taken is taken into account in the same way as in 
\cite{IJL,BI}. For small values of $Q^2$ the form factors are dominated by the 
meson dynamics, whereas for large values the modification from dimensional 
counting laws from p-QCD can be taken into account by scaling 
$Q^2$ with the strong coupling constant \cite{gari}. 

The Dirac and Pauli form factors that correspond to the strange current 
are factorized in terms of the product of an intrinsic part $g(Q^2)$ and 
a contribution from the meson cloud as 
\ba
F_{1}^{s}(Q^{2}) &=& \frac{1}{2}g(Q^{2})\left[ 
\beta_{\omega}^s \frac{m_{\omega}^{2}}{m_{\omega }^{2}+Q^{2}} 
+\beta_{\phi}^s \frac{m_{\phi}^{2}}{m_{\phi }^{2}+Q^{2}}\right] ~, 
\nonumber\\
F_{2}^{s}(Q^{2}) &=& \frac{1}{2}g(Q^{2})\left[ 
\alpha_{\omega}^s \frac{m_{\omega}^{2}}{m_{\omega }^{2}+Q^{2}}
+\alpha_{\phi}^s \frac{m_{\phi}^{2}}{m_{\phi }^{2}+Q^{2}}\right] ~.
\label{sff}
\ea

The $\beta$'s and $\alpha$'s in Eqs.~(\ref{ff}) and (\ref{sff}) 
are not independent of one another. The coefficients appearing in the 
isoscalar and strange form factors depend on the same nucleon-meson and 
current-meson couplings \cite{Jaffe}. In addition, they are constrained 
by the electric charges and magnetic moments of the nucleon 
\ba
\alpha_{\omega} &=& \mu_p + \mu_n -1 - \alpha_{\phi} ~,
\nonumber\\
\beta_{\omega} &=& - \beta_{\phi} \tan(\Theta_0+\epsilon)/\tan \epsilon ~.
\label{coef1}
\ea
The strange couplings can be expressed as 
\ba
\beta_{\omega}^s/\beta_{\omega} =  
\alpha_{\omega}^s/\alpha_{\omega} &=& 
-\sqrt{6} \, \sin \epsilon/\sin(\Theta_0+\epsilon) ~,
\nonumber\\
\beta_{\phi}^s/\beta_{\phi} =  
\alpha_{\phi}^s/\alpha_{\phi} &=& 
-\sqrt{6} \, \cos \epsilon/\cos(\Theta_0+\epsilon) ~.  
\label{coef2}
\ea
Eqs.~(\ref{coef1}) and (\ref{coef2}) only depend on the mixing angles 
between the $\omega$ and $\phi$ isoscalar mesons 
\ba
\left| \omega \right> &=& \cos \epsilon \left| \omega_0 \right> 
- \sin \epsilon \left| \phi_0 \right> ~,
\nonumber\\
\left| \phi \right> &=& \sin \epsilon \left| \omega_0 \right> 
+ \cos \epsilon \left| \phi_0 \right> ~,
\ea
where $\left| \omega_0 \right>=\left( u \bar{u} + d \bar{d} \right)/\sqrt{2}$ 
and $\left| \phi_0 \right> = s \bar{s}$ are the ideally mixed states 
characterized by the mixing angle $\theta_0$ with 
$\tan \theta_0 = 1/\sqrt{2}$. The mixing angle $\epsilon$ has been determined from 
the decay properties of the $\omega$ and $\phi$ mesons to be $\epsilon=0.053$ rad  
\cite{Jain}. 

\section{Results}

In order to calculate the nucleon form factors in the two-component model the 
five coefficients, $\gamma$ from the intrinsic form factor, $\beta_{\phi}$ and 
$\alpha_{\phi}$ from the isoscalar couplings, and $\beta_{\rho}$ and  
$\alpha_{\rho}$ from the isovector couplings, are determined in a least-square 
fit to the electromagnetic form factors of the proton and the neutron using the 
same data set as in \cite{BI}.  

\begin{figure}[htb]
\vspace{-0.8cm}
\insertplot{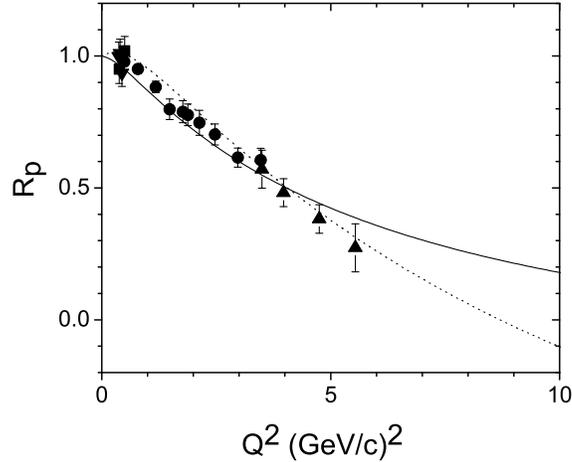}
\vspace{-1.2cm}
\caption[]{Comparison between the experimental and theoretical proton 
form factor ratio $R_p=\mu _{p}G_{E_{p}}/G_{M_{p}}$. 
The experimental data are taken from a 
compilation in \cite{BI}. The solid line is from the 
present calculation and the dotted line from \protect\cite{IJL}.}
\label{rp}
\end{figure}

\begin{figure}[htb]
\vspace{-0.8cm}
\insertplot{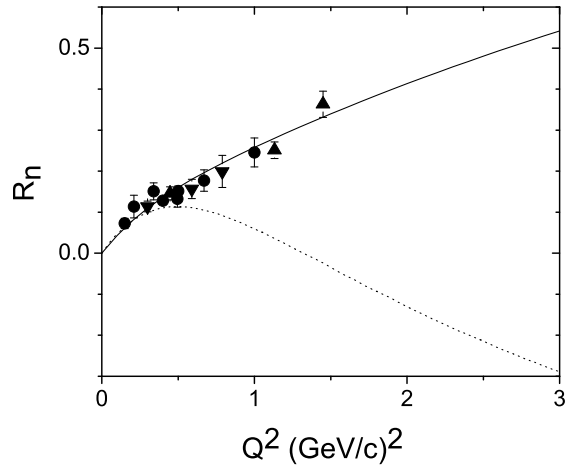}
\vspace{-1.2cm}
\caption[]{As Fig.~\ref{rp}, but for the neutron 
form factor ratio $R_n=\mu _{n}G_{E_{n}}/G_{M_{n}}$.}
\label{rn}
\end{figure}

In Figs.~\ref{rp} and \ref{rn}, the form factors ratios for the proton 
and neutron are compared with experimental data. The linear drop in 
the proton form factor ratio was predicted as early as 1973 in a VMD 
model \cite{IJL} (dotted line) and later also in a chiral soliton model 
\cite{soliton}. The experimental data for the neutron form factor ratio 
\cite{madey} are in agreement with the VMD model of \cite{IJL} for small values 
of $Q^{2}$, but not so for higher values of $Q^2$. The present calculation 
(solid line) is in good agreement with the data, especially for the neutron. 

The strange form factors can now be obtained by combining Eqs.~(\ref{sff}) 
and (\ref{coef2}). As a result, the strange magnetic moment is given by
\ba
\mu_s = G_M^s(0) = \frac{1}{2} (\alpha_{\omega}^s+\alpha_{\phi}^{s}) 
= 0.315 \, \mu_N ~,
\ea
a positive value, in contradiction to most theoretical values, but in 
agreement with recent experimental evidence from the SAMPLE Collaboration 
which determined the strange magnetic form factor at $Q^2=0.1$ (GeV/c)$^2$ 
to be $G_M^s=0.37 \pm 0.20 \pm 0.26 \pm 0.07$ \cite{Spayde}. An analysis 
of the world data gives $G_M^s(0.1)=0.55 \pm 0.28$ \cite{Aniol05b}. 

\begin{figure}[htb]
\vspace{-0.8cm}
\insertplot{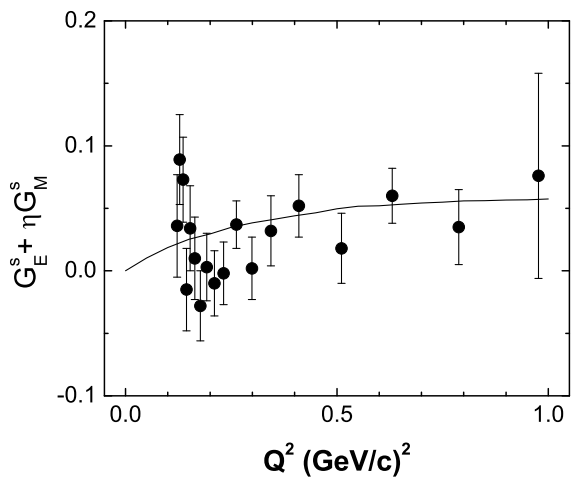}
\vspace{-1.2cm}
\caption[]{Comparison between theoretical and experimental values of 
the strange form factor $G_E^s + \eta G_M^s$.  
The experimental values are taken from the G0 Collaboration 
\cite{Armstrong}.} 
\label{G0}
\end{figure}

Fig.~\ref{G0} shows that the calculated values for the strange form factor 
combination $G_E^s + \eta G_M^s$ are in good agreement with the experimental 
ones obtained recently by the G0 Collaboration \cite{Armstrong}.   

\section{Summary and conclusions}

In this contribution, the recent data on electromagnetic and weak form 
factors of the nucleon were analyzed in a two-component model which consists 
of an intrinsic (three-quark) structure and a meson cloud whose effects were 
taken into account via VMD couplings. The parameters in the model are 
completely determined by the electromagnetic form factors of the proton and 
neutron. The strange couplings follow directly from the electromagnetic ones 
and do not involve any new parameters. On the contrary, the fact that the 
net contribution of the strange quarks to the electric charge of the nucleon 
is zero, leads to an extra constraint relating $\beta_{\omega}$ and $\beta_{\phi}$. 

A good overall agreement is found both for the form factor ratios of the 
proton and neutron and the strange form factor as measured by the G0 
Collaboration. The size of the intrinsic structure is found to be $\sim 0.49$ fm.  
The strange magnetic moment is calculated to be positive 
in agreement with recent data from parity violation electron scattering. 

The first results from the SAMPLE \cite{Spayde}, HAPPEX 
\cite{Aniol05b,Aniol,Aniol05a}, G0 \cite{Armstrong} 
and PVA4 \cite{Maas} collaborations have 
shown evidence for a nonvanishing strange quark contribution to the charge 
and magnetization distributions of the nucleon. Future experiments hold 
great promise to be able to unravel the contributions of the different 
quark flavors to the electromagnetic and axial form factors, and thus to 
give new insight into the complex internal structure of the nucleon. 

\section*{Acknowledgments}
It is a pleasure to thank F. Iachello and K. de Jager for many interesting 
discussions. This work was supported in part by a grant from CONACYT, Mexico.

\vfill\eject
\end{document}